\documentclass[10pt,a4]{article}

\usepackage[top=0.85in]{geometry}
\usepackage{amsmath,amssymb}
\usepackage{changepage}
\usepackage{textcomp,marvosym}
\usepackage{cite}
\usepackage{nameref}
\usepackage[pdfborder={0 0 0}]{hyperref}
\usepackage[nopatch=eqnum]{microtype}
\usepackage[table]{xcolor}
\usepackage{array}
\usepackage{graphicx}

\DisableLigatures[f]{encoding = *, family = * }

\newcolumntype{+}{!{\vrule width 2pt}}

\newlength\savedwidth

\usepackage[aboveskip=1pt,labelfont=bf,labelsep=period,justification=raggedright,singlelinecheck=off]{caption}

\makeatletter
\renewcommand{\@biblabel}[1]{\quad#1.}
\makeatother

\begin{document}
\vspace*{0.2in}

\begin{flushleft}
{\Large\textbf\newline{The Ontoverse: Democratising Access to Knowledge Graph-based Data Through a Cartographic Interface }}
\newline


Johannes Zimmermann\textsuperscript{1\Yinyang*},
Dariusz Wiktorek\textsuperscript{1,2},
Thomas Meusburger\textsuperscript{1},
Miquel Monge-Dalmau\textsuperscript{3},
Antonio Fabregat\textsuperscript{4},
Alexander Jarasch \textsuperscript{5}, 
Günter Schmidt\textsuperscript{1},
Jorge S. Reis-Filho\textsuperscript{6},
T. Ian Simpson\textsuperscript{7\Yinyang}
\\
\bigskip
\textbf{1} Computational Pathology GmbH, AstraZeneca, Munich, Germany
\\
\textbf{2} ADLA IT Ltd., Hurlingham Studios, London, UK
\\
\textbf{3} Knowledge Graph Team, Enterprise Data Office, AstraZeneca, Barcelona, Spain
\\
\textbf{4} Knowledge Graph Team, Enterprise Data Office, AstraZeneca, Cambridge, UK
\\
\textbf{5} Neo4j Germany GmbH, Life Sciences Dept., Munich, Germany
\\
\textbf{6} Cancer Biomarker Development, Oncology R\&D, AstraZeneca, Gaithersburg, MD, USA
\\
\textbf{7} School of Informatics, The University of Edinburgh, Edinburgh, UK
\\
\bigskip

%
%
\Yinyang These authors contributed equally to this work.

*johannes.zimmermann@astrazeneca.com

\end{flushleft}

\section*{Abstract}
As the number of scientific publications and preprints is growing exponentially, several attempts have been made to navigate the sheer volume of this complex and increasingly detailed landscape. These have almost exclusively taken unsupervised approaches that fail to incorporate domain knowledge. As a consequence, these emerging landscapes lack the structural organisation required for intuitive interactive human exploration and discovery. Especially in highly interdisciplinary fields, a deep understanding of the connectedness of research works across topics is essential for generating insights. We have developed a unique approach to data navigation that leans on geographical visualisation and uses hierarchically structured domain knowledge to enable end-users to explore knowledge spaces grounded in their desired domains of interest. This can take advantage of existing ontologies, proprietary intelligence schemata, or be directly derived from the underlying data through hierarchical topic modelling. Our approach uses natural language processing techniques to extract named entities from the underlying data and normalise them against relevant domain references and navigational structures. The knowledge is integrated by first calculating similarities between entities based on their shared extracted feature space and then by alignment to the navigational structures. The result is a knowledge graph that allows for full text and semantic graph query and structured topic driven navigation. This allows end-users to identify entities relevant to their needs and access extensive graph analytics. The user interface facilitates graphical interaction with the underlying knowledge graph and mimics a cartographic map to maximise ease of use and widen adoption. 
We demonstrate an exemplar project using our generalisable and scalable infrastructure for an academic biomedical literature corpus that is grounded against hundreds of different named domain entities.

\section*{Introduction}
Scientific articles are being published at an exponential rate with one recent study reporting an increase of roughly 50\% across the period 2016-2022 at the end of which a staggering 3 million articles were published \cite{hanson_strain_2023}. As a result, assessing and navigating emerging scientific knowledge effectively have become increasingly challenging for researchers. While search engines and natural language processing (NLP)-based methods can be used to enable targeted retrieval of relevant publications, there is a broader need for a holistic approach to explore scientific knowledge \cite{Mendoza_Agarwal_Blackshaw_Bol_Fazzi_Fiorini_Foreman_George_Johnson_Martin_etal._2024}; especially in the context of drug and biomarker development, at the intersection of chemistry, biology, medicine and advanced artificial intelligence.

A number of approaches have been proposed to represent biomedical literature using articles indexed in PubMed \cite{González-Márquez_Schmidt_Schmidt_Berens_Kobak_2024} or pre-prints listed in arXiv \cite{knegjensPaperscape2013}. These are typically implemented as networks in which the similarity between publications (nodes) is calculated and relationships (edges) created where a similarity threshold has been met. It is common to use dimensionality reduction techniques calculated from paper feature space to embed the spatial relationships between nodes in two dimensions in a purely data-driven manner. Whilst this can indeed successfully reveal clusters of publications that are organised by topic, such techniques neither capture true topological relationships nor map them consistently onto structured domain knowledge. These approaches lack basic paper meta-data and are rarely enriched with augmented data derived from textual analysis. Together these two classes of paper feature are necessary to enable advanced analytical, navigational, and user-interface functions. A current axiomatic assumption in the generation of literature graphs is that publications are unique nodes. This naturally precludes the possibility of an article's influence being felt in more than one position in the graph space. This precludes the ability to observe a paper's influence in multiple topic areas grossly over-simplifying it's place within the knowledge space meaning that a true understanding of it's importance is lost. Since a unifying feature of all scientific publications is that they describe and connect to many different concepts, this should be reflected in any system that aims to visualise their position and relationships.

We propose a novel approach to model and visualise knowledge spaces in general and illustrate this by implementing our paradigm in the biomedical literature domain. Our approach aims to optimise both the data modelling and user interaction aspects jointly. We adopt a visual approach analogous to that used for Google Maps in which our knowledge domain is presented as a cartographic map that users explore by panning and zooming. The map provides search functionality and is annotated with useful sign-posts (labels) and paths (edges) that capture the organisation and connectivity between entities and topics. The topic structure can be user provided or derived from any hierarchical graph such as an ontology. 

The best way to structure knowledge contained within a literature corpus is an open research question. An exemplar use case we are presenting is a library of scientific publications aimed to provide one coherent knowledge base to a community with a focus on computational pathology for discovery of predictive biomarkers in oncology. The repository spans domains from clinical oncology to AI science and all application fields in between, targeting highly diverse audiences in this inherently interdisciplinary field, with their specific mindsets, paradigms and jargons. A core interest of the knowledge base is to ensure a common understanding between the different "tribes" and increase productivity in Oncology R\&D by providing state-of-the-art, rigorously selected, relevant information. Its implementation in a classical reference management system, Zotero \cite{Zotero2016}, though, did not encourage spontaneous knowledge discovery workflows and the membership-based closed setup hindered broader adoption, which motivated the development of an interface overcoming these constraints. Gradually, the transferability of the described approach to other, unrelated use cases became obvious, and we evolved the technical framework of the specific application into a generic platform.

\begin{figure}[!ht]
    \centering
    \includegraphics[width=1\linewidth]{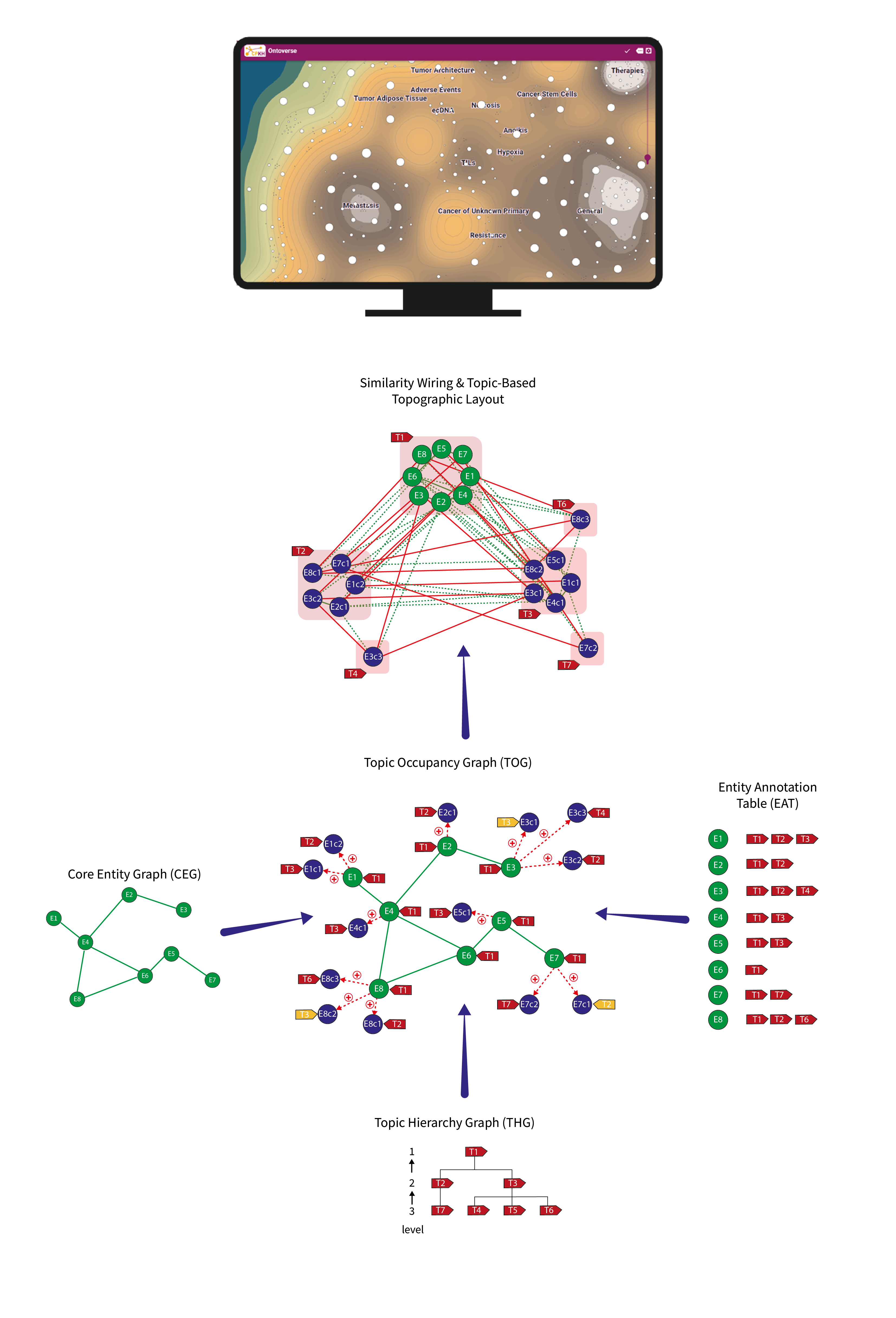}
    \caption{{\bf{{\textbf{Ontoverse}}:}} A cartographic user interface nurtured by an ensemble of three Knowledge Graphs, allowing for the seamless navigation of a knowledge landscape.}
    \label{fig1:graphical-abstract}
\end{figure}

There are several modelling and natural processing technologies that have the potential to allow for rich, flexible, and generalisable annotation of information from text documents. We refer to the text objects in this project as publications, though the methods could readily be extended to other data types. Our system comprises three distinct components i) a data model that incorporates rich paper features and is implemented in a knowledge graph architecture developed in Neo4J \cite{besta_demystifying_2023}, ii) a natural language processing pipeline that extracts concepts from text guided by domain ontologies, and iii) a state-of-the-art graphical user interface to visualise and interact with the underlying graph through a web browser (\hyperref[fig1:graphical-abstract]{\textbf{Fig. 1}}).

\clearpage

\clearpage

\section*{Requirements Analysis}
\subsection*{Core Design Principles}

\textbf{Domain flexibility:} From the outset and by design models are generalisable so that they can in principle be applied to any text corpus and used for any domain where a reference ontology or organising dictionary is available. Additionally, in situations where a purely data driven navigational structure is appropriate, for example where no such reference is available, the hierarchical topic modelling procedure will build a navigational structure \textit{ab initio}. 

\noindent{\textbf{Topic expressivity}: The model allows for a unique node (in the discussed use case the entities are publications) to be visualised multiple times in the geographic representation of the network.}

\noindent{\textbf{Normalisation:} The model maps identified elements to selected domain ontologies and/or vocabularies}






\subsection*{Human Computer Interaction}
\textbf{Intuitive use:} The navigational structure is topic driven and hierarchical to facilitate granular knowledge domain exploration, which suggests a geographical approach for visualising the data. Employing elements from cartographic design like altitude lines and a feature association-based colour scheme, topography rendering is used as metaphor to give bodies of knowledge a tangible dimension and help users intuitively grasp topic structure and significance. For an entity in focus, indicator lines guide the user to all topics and associated geographical regions related to this entity.

\noindent{\textbf{Graph query and targeted navigation:} Individual entities as well as topics on any level of granularity are directly accessible using intuitive text search functionality, and guide the user to the respective geographic location.}

\noindent{\textbf{Communication of results:} Entities of interest are downloadable as lists for generating reading recommendations, bibliographical references or the like. Regions of the landscape representing topics and subtopics are shared as URL comprising geographical coordinates (longitude, latitude, and altitude representing the zoom level).}

\section*{Implementation}
\subsection*{Data Acquisition and Processing}

We have designed and implemented a data ingestion framework for a curated Zotero database of c.7800 entries that has been tagged and organised into hierarchical topic folders using a controlled vocabulary.



When papers are imported into Zotero, the platform attempts to extract critical publication meta-data either from a record (such as a PubMed record) or from the features of a PDF document. There are inconsistencies in this process that make it a sub-optimal long-term solution for downstream tasks that require high fidelity. For project development purposes, this approach has several advantages, namely that it; i) naturally allows for simple user-driven bespoke collection creation, ii) has an organisational structure that can be manually curated, and iii) is based on a sqlite relational database back-end that is trivially programmatically accessible.

We automatically extracted bibliographical metadata from Zotero into a publication instance that was populated with attributes as well as the topic(s) the publication had been assigned to. Critically the objects included the title and abstract that are used as substrate for downstream natural language processing tasks.

To perform Named-Entity Recognition (NER), we developed a biomedical NLP pipeline using the Allen Institute ScispaCy biomedical text processing environment which employs an ensemble of 5 state-of-the-art pre-trained models to identify biomedical entities within our input text: sciBERT, bioNLP13cg, bc5cdr, jnlpba, and craft. Text was first pre-processed, tokenized and then input to each model. The resulting entities were subsequently normalised to unique concept identifiers in the Unified Medical Language System (UMLS). Publication objects were then annotated with the unique concept identifiers to be used later in i) linking publications as explicit relationships in the database to the topic hierarchy graph, ii) topic modelling, and iii) calculation of publication similarity metrics. Annotations were filtered by selected sources from the
UMLS MetaThesaurus including the Human Phenotype Ontology (HPO), DRUGBANK, Medical Subject Headings (MeSH), Human Genome Nomenclature Committee (HGNC), National Cancer Institute (NCI), and RXNORM.

\subsection*{Knowledge Graph Implementation}


Implementing a knowledge graph in Neo4j is highly beneficial due to the inherently heterogeneous and highly connected nature of biomedical data. Neo4j stores data natively in a connected way, maintaining its network structure, which is ideal for representing complex relationships. 
Native \textbf{graph processing} refers to how a native graph database processes database operations, including both storage and queries. A key capability of a native graph database is the ability to navigate through the connections in the data quickly – without the overhead of index lookups or other join strategies. This capability to traverse the related data without the overhead of an index lookup for each move across a relationship is called \textit{index-free adjacency}.
Non-native graph processing often needs to use index lookups to get to the next element in a chain for completing a read or write transaction; this may be okay with just one or two relationships, but in today’s use cases with hundreds or thousands, it won’t work. The complexity of an index lookup is often O(log(n)) versus the O(1) for following a direct pointer of the relationship to a target node.
 The flexibility of Neo4j's data schema is perfectly suited for biomedical data, making it easy to model, understand, and extend according to specific needs and use cases. As a native graph database, Neo4j allows entities to be effortlessly connected to additional layers such as ontology terms, enhancing data integration and analysis. Neo4j's scalability ensures that the graph can be extended with more data and adapted for future use cases. We selected Neo4J as our graph database infrastructure as it also provides powerful graph analytics capabilities that will be used to expand functionality within the Ontoverse system. This will include integration of advanced NLP, query, navigation, and user-interaction features.

A series of python scripts generates the knowledge graph architecture, comprising: 

\begin{itemize}
\item Core Entity Graph (CEG) - the master structure with richly annotated nodes (representing entities) and edges (representing similarities between entities)
\item Topic Hierarchy Graph (THG) - the navigational backbone 
\item Topic Occupancy Graph (TOG) - the feature that enables multi-topic occupancy for entities.  
\end{itemize}

\subsubsection*{Connection: Core Entity Graph (CEG)}
To create the Core Entity Graph (\hyperref[fig2:graph-essentials]{\textbf{Fig. 2A}}) consisting of publication nodes and edges connecting similar entities, we first explored a selection of metrics for calculating paper similarity. The most efficient and stable metric we have currently employed consists in counting the number of concepts shared between publications (remembering that this covers thousands of unique concept terms) and was used to create edges connecting entities with an associated count weighting captured as a relationship property. Albeit conservative, this approach provides the foundation for future developments incorporating  more complex semantic information in the metric calculation to quantify similarity between non-identical terms more sensitively.

To attain a base network with an informative density we sparsify the network using a threshold on edge weights to only include edges connecting publications with at least 5 concepts in common. In practice, this maintains a single connected component (no orphan entities) without obscuring the inherent community structure of the graph. In future work we will explore the relative benefit of using alternate network sparsification approaches such as k-Nearest-Neighbour and minimum spanning tree based algorithms.

We finally injected our Networkx CEG object directly into a Neo4J graph instance alongside a rich set of node and edge features to facilitate flexible later interrogation and analysis.

\subsubsection*{Navigation: Topic Hierarchy Graph (THG)}

\textbf{Manual THG Generation (mTHG):} To generate the navigational structure we built a hierarchical graph in Networkx \cite{Hagberg_Schult_Swart_2008} that captured the topic folder structure of the underlying Zotero database (\hyperref[fig2:graph-essentials]{\textbf{Fig. 2B}}).  

Before connecting this THG to the underlying CEG, we first had to address a critical feature of navigational trees which have transitive properties. By definition, superior folders in the ZoteroDB are broader topic terms and their sub-folders more specific sub-topics. That means that if an entity is put into a sub-folder it is also a member of the parental topic. To correctly capture this transitive property, which is also shared in ontologies, algorithms to back-propagate topic annotations towards the parental root nodes were written to ensure that entities were annotated to every topic on the path(s) from their explicit annotation upwards. We updated entity node features to hold both the uniquely annotated topic and the topic path back to root. 

Finally, to connect the THG to the CEG we used the Neo4J Cypher query language to create relationships (Neo4J terminology for edges) between entities and all topic nodes in the THG where the unique topic ID appeared in the entity topic feature element. This hardwired together the two graphs creating the initial architecture for hierarchical navigation.

\textbf{Data Driven THG Generation (dTHG):} An important use case to accommodate is one where there is no known navigational structure or where there is a requirement to allow the entity data itself to generate a topic structure to layout the topology in an unbiased way. To facilitate dynamic topic modeling, we implemented - using the Python corex package \cite{Gallagher_Reing_Kale_Steeg_2017} - a data driven approach to THG generation that uses correlation explanation to discover collections of words that aggregate into discernible topics. We used the concept vectors previously identified through our ScispaCy pipeline as input to corex topic modelling, specifying the number of topics to be discovered at each level for however many levels (navigational depth) included in the final topographical model. Depths of c.5 levels with a pyramid number of topics ranging from c.200 down to c.10 at the top-level were found to give a generally tractable and intuitive granularity. In future work this could be explored in more detail, but is inherently customisable by design. 

This initial approach allows the assignment of entities to topics at each of the levels, but does not connect the levels to each other. There are two main ways this can be done i) use topics from a lower layer as input to the superior layer and ii) to map topics exhaustively by similarity between layers. Both approaches were explored; and and the latter found to be more accurate and interpretable. In this approach, the pairwise similarities between entities assigned to topics between layers were calculated and utilized to create a THG edge between the highest scoring pair of each layer. This naturally builds the fully connected data-driven hierarchical network that we need. 

\subsubsection*{Distribution: Topic Occupancy Graph (TOG)}

The final methodological development included in the requirements for the system was extremely complex to implement as it required that any given entity could be observable in the graph in multiple locations. As illustrated in the development process described so far, the underlying CEG is comprised of unique publication nodes. This is crucial for many reasons not least that analytic algorithms that have been developed for the analysis of network structures depend on this axiomatically. That includes everything from basic clustering algorithms to advanced machine learning approaches such as embedding and inference with graph neural networks, and emerging approaches to learn explicitly on knowledge graph structures. 

\begin{figure}[!ht]
    \centering
    \includegraphics[width=1\linewidth]{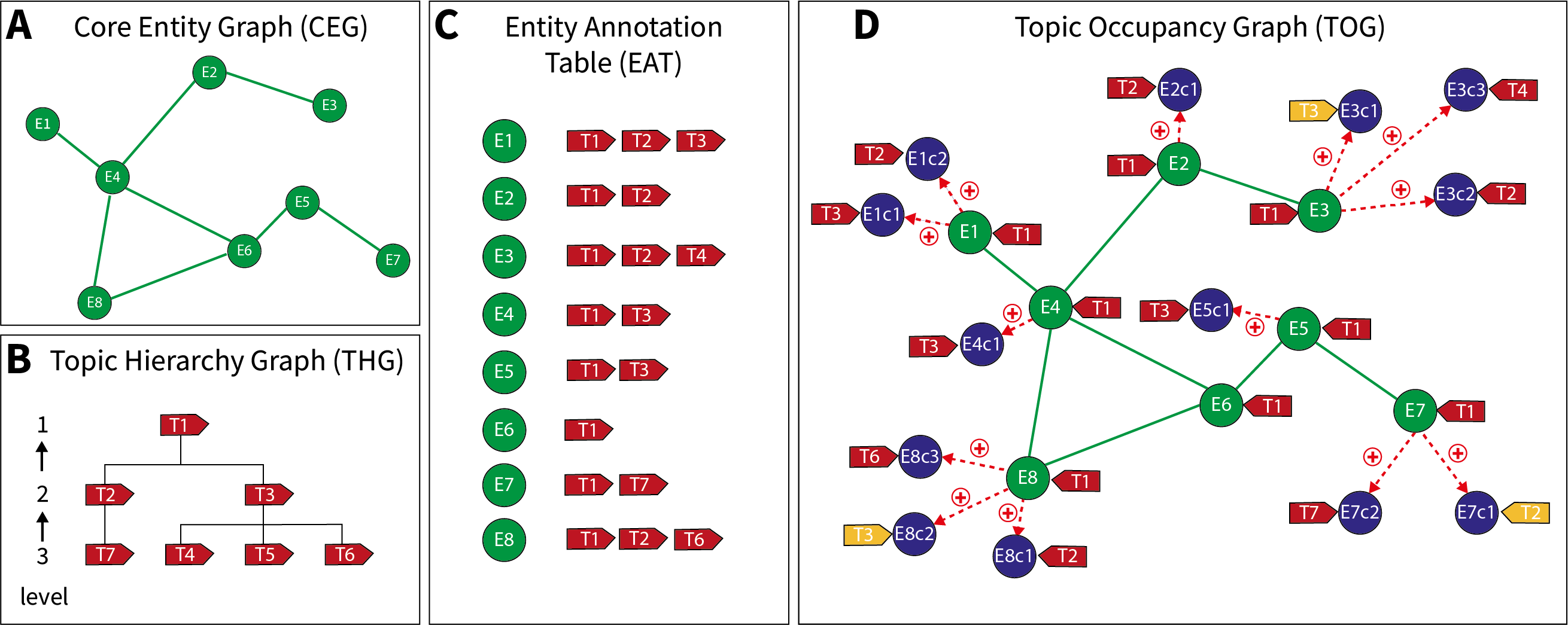}
    \caption{{\bf{Overview of the core {\textit{Ontoverse}} building blocks.}} The Core Entity Graph (CEG) is comprised of entity nodes, in the use case we are discussing here publications (green circles), and edges (green lines) representing inter-entity similarity ({\bf{A}}). The CEG is anchored by alignment with a topic hierarchy graph (THG) that can be manually or dynamically generated by topic modelling in the graph domain space, or derived from preexisting domain ontologies ({\bf{B}}). The Entity Annotation Table (EAT) defines links between topics and entities, it can be generated by manual curation or performing named-entity-recognition (NER) on entity meta-data ({\bf{C}}). Entities that are associated with more than one topic are represented by entity clones (blue nodes) that are dynamically created based on their pathway relationships within the THG. The entity clone concept allows spatially separate topics whilst maintaining the topology of the graph. We induce topic association (yellow topic tags) to enable logically coherent and seamless navigation guided by the THG ({\bf{D}}).}
    \label{fig2:graph-essentials}
\end{figure}

\begin{figure}[!ht]
    \centering
    \includegraphics[width=1\linewidth]{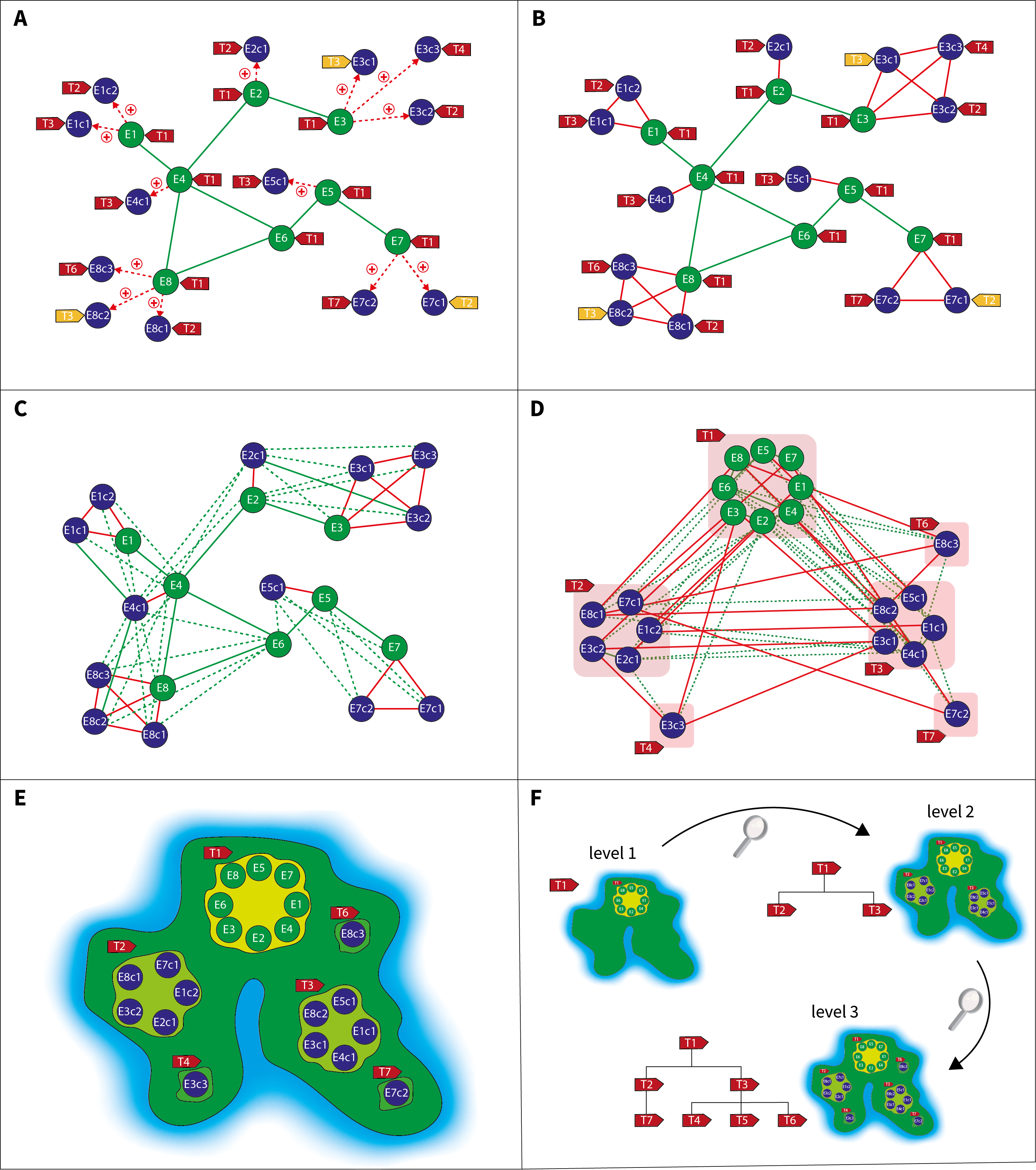}
    \caption{{\bf{Graph Connectivity \& Topographic Navigation}}. Parental entities (\textit{green circles}) are connected if they are annotated to the same topic (\textit{solid green lines}) then entity clones (\textit{blue circles}) are generated based on their mappings to topics either by being directly annotated to the topic (\textit{red flags}) or by induction (\textit{gold flags}) the latter topics falling on the transitive path between the parent entity topic and a sub-topic in the THG (\textbf{A}). Matching edges are created to connect all parental entities with their clones (\textit{red lines})(\textbf{B}). We next generate all necessary edges between entity clones to satisfy the parental mappings, for example if E1$\leftrightarrow$E4 then \{E1c1$\leftrightarrow$E4, E1c2$\leftrightarrow$E4, ...\}. Edges are expanded to incorporate the connectivity of entity clones with "within-topic" (\textit{solid green lines}) or "between-topic" (\textit{dashed green lines}) edges to allow exploration of their proximal and distal topic environment (\textbf{C}). Graph topology is then re-organised using a nested radial algorithm to arrange topics and their sub-topics proximal to each other (\textbf{D}). We visualise the graph as a cartographic map with contour lines and colouration reflecting entity density; densely packed locations as hills, sparse areas as valleys and the "\textit{sea}" separating areas that are poorly connected (\textbf{E}). We can view the map at different levels of granularity by selecting the THG depth (\textbf{F}). All edges are weighted based on the calculated similarity score between the corresponding parental entities.}
    \label{fig3:graph-design}
\end{figure}

We adopted a novel approach, the Topic Occupancy Graph (TOG; \hyperref[fig2:graph-essentials]{\textbf{Fig. 2D}}) in which entity clones are created in the CEG to allow for multi-topic occupancy (positioning entities in more than one location in the landscape) and enable additional downstream analytic options. This was achieved by writing a series of algorithms to generate a topic path structure for every entity and spawning novel entity clones as we navigate up the path to the topic root node (\hyperref[fig3:graph-design]{\textbf{Fig. 3}}). To illustrate with the simplest example: If an entity is annotated to 1 topic in level 1 of the THG (the highest level) and to 2 topics in level 2 of the THG then we retain the core entity node in level 1 (the original CEG), but spawn two clones of the entity in level2. Critically these entity clones have unique identifiers and type allowing to identify them uniquely and tag them with only a single topic ID.

\clearpage

\subsection*{User Interface Implementation}

\subsubsection*{Back-end service}

\begin{figure}[!ht]
    \centering
    \includegraphics[width=1\linewidth]{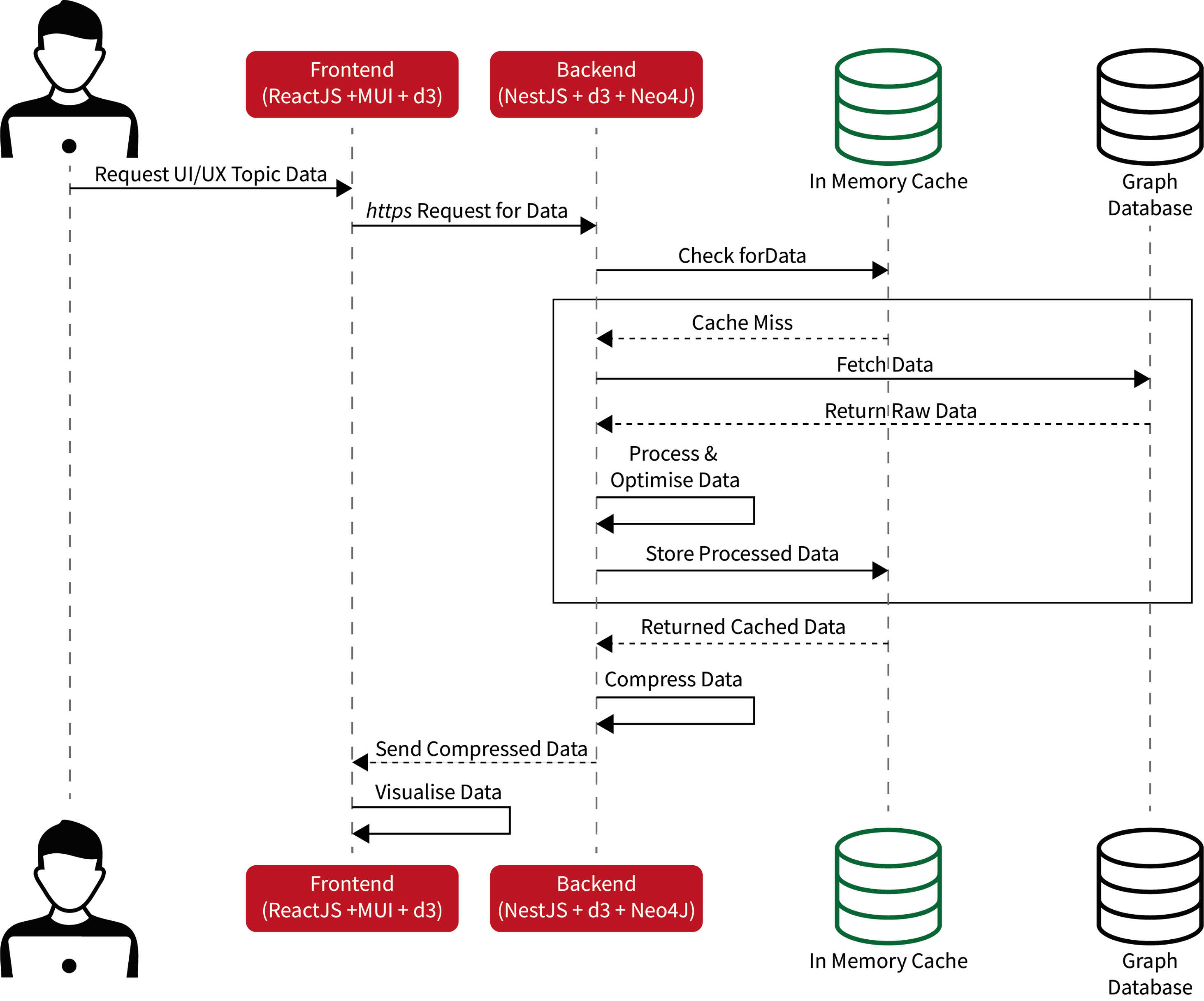}
    \caption{{\bf{Sequence Diagram of the Ontoverse flow}} within the layers of a technical stack.}
    \label{fig4:enter-label}
\end{figure}

The back-end architecture was designed to maximise the performance of the front-end visualisation of the Ontoverse map to provide the most intuitive visual representation of the underlying data (\hyperref[fig4:enter-label]{\textbf{Fig. 4}}). Therefore, the back-end service focuses mostly on data retrieval, optimisation, and pre-processing with the help of  NestJS, d3 and Neo4j libraries. The NestJS framework enforces best practice for REST API development and establishes the template for implementing existing features and allows for addition of future functionalities with ease by adding new endpoints, utilising NestJS design patterns. 

In general, the  API service of the main end-point, responsible for obtaining the optimised list of the entities, executes the following process: It fetches the data from the Knowledge Graph database, optimises the data structure (edges list, entity list, topic-leaves hierarchy, etc) and its properties' references. Next, it pre-calculates essential visualisation-related variables, such as node positions. This processed data is then stored in an in-memory cache for rapid access. When a HTTPS request is made by the frontend upon user interaction, the backend retrieves the data from this cache (if available), compresses it to reduce its size significantly, and then sends it to the front-end client. 

This streamlined process ensures quick response times and minimises data transfer overhead. Future versions will explore further optimisations and introduce more specific API endpoints, enhancing adaptability and functionality.

\textbf{}

\subsubsection*{Front-end web application}

The front-end utilises various popular tools that enhance both functionality and design, supporting the advanced data visualisation of the Ontoverse map. They also facilitate the development of efficient user interfaces with a modern and uniform aesthetic across all components, setting the standards for future UI of the anticipated new functionalities.

The application displays a layered map with the user interface that form a cohesive view of an exploratory tool of the Ontoverse. The topmost layer is the UI layer, which includes the application's top bar and a side panel with various interactive widgets. Beneath this, the Ontoverse map layer is structured into several detailed sub-layers. 

The top sub-layer features entity nodes that reveal their descriptions when sufficiently zoomed in, with topic labels that change their visibility at different zoom levels to provide contextual insights.

The next sub-layer holds the contour lines with gradient colours illustrating the topographical features of hills and valleys. This map layer is set against a blue background representing waters, visually separating the main topics and creating distinct "continents" on the map.

\subsubsection*{Geography}

\begin{figure}[!ht]
    \centering
    \includegraphics[width=1\linewidth]{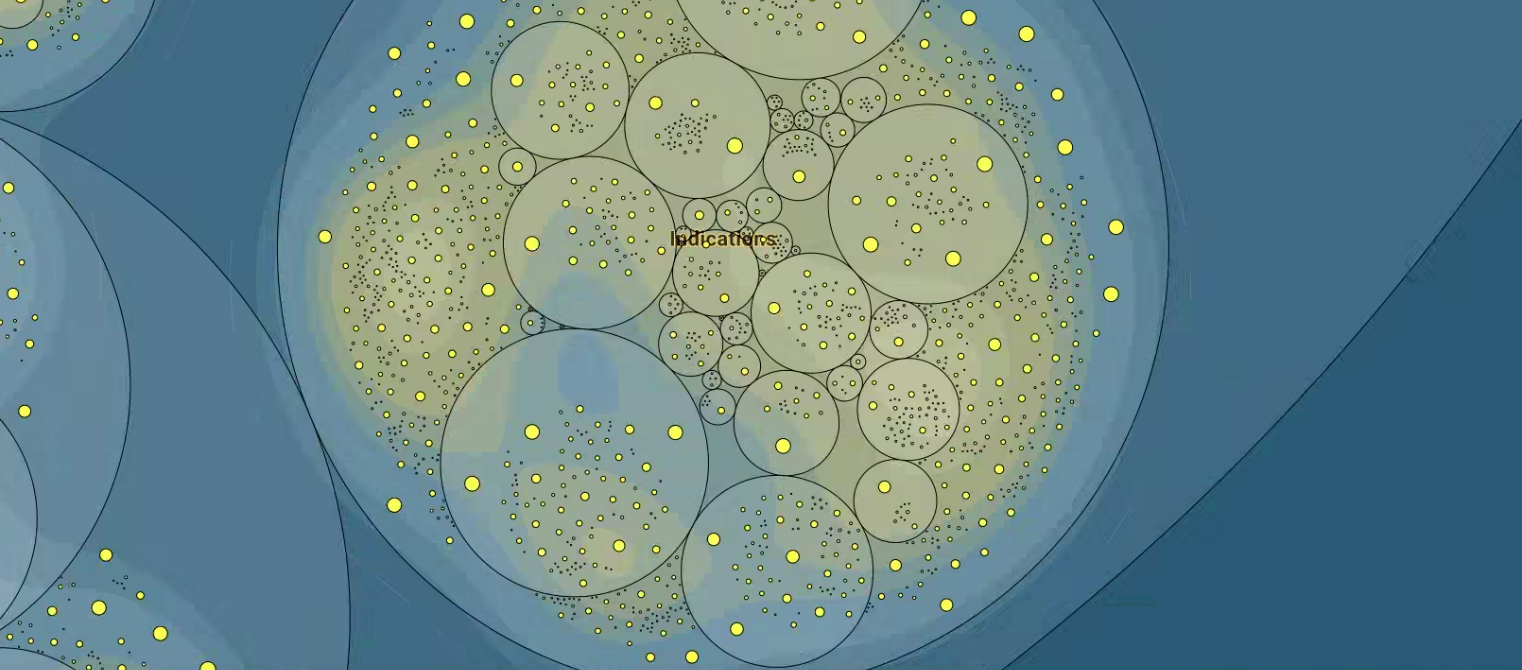}
    \caption{{\bf{Spatial layout of entities in the graph}} as dictated by circular packing.}
    \label{fig5:circle-packing}
\end{figure}

Internally, topics are spatially arranged by means of circle packing layout \cite{Collins_Stephenson_2003}, \cite{kennethstephensonCirclePackingMathematical2003}, which provides a direct visual translation of the topic hierarchy into its geographical representation and ensures excellent readability of the map. The algorithm arranges inferior topics as smaller circles within larger circles representing superior topics based on the hierarchical structure of the topics  (\hyperref[fig5:circle-packing]{\textbf{Fig. 5}}), making it visually intuitive. Earlier attempts to employ force-directed graphs resulted in extremely busy clusters of points, which hampered map comprehensibility. In contrast, a circle packing arrangement helps to distinguish the different hierarchical levels and provides a clear understanding of each topic's relative size and scale. 

Additionally, the topographic elevation, determined by the nested-ness and size of the topics, is visually represented through contour lines. These lines feature a colour gradient that spans from a minimum (sea level) to various maxima (mountain tops) where the criteria are highest, effectively illustrating changes in altitude.

\textbf{}

\subsubsection*{Interaction}

The interface is designed to provide the user with an experience matching the current state-of-the-art in online map tools. Two major entry points exist: Exploratory navigation is encouraged through key elements like panning (to soar over topics) and zooming (to change granularity of display), whereas directed navigation is facilitated by search and query functionality. 

\textbf{}

\begin{figure}[!ht]
    \centering
    \includegraphics[width=1\linewidth]{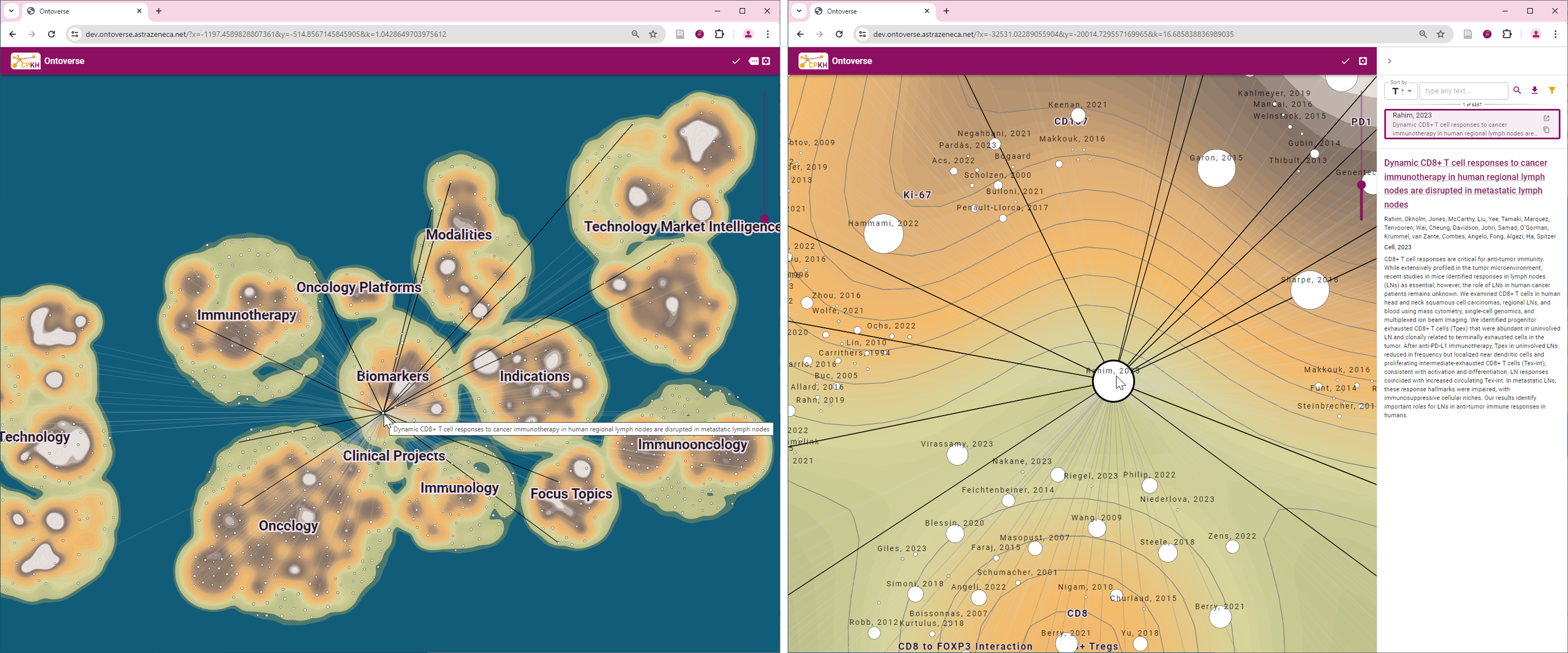}
    \caption{{\bf{\textit{Ontoverse} user interface with core functionality for exploratory navigation.}} Left panel: Bird's eye view at high altitude. Right panel: Granular view at low altitude. In both cases, the same paper with significant multi-occupancy across various topics is selected. }
    \label{fig6:exploratory-navigation}
\end{figure}

\textbf{Exploratory navigation: }
To achieve an initial understanding of a hitherto novel domain of knowledge, the user can stroll across the landscape by panning on a high “cruising altitude” over the meta-domains (\hyperref[fig6:exploratory-navigation]{\textbf{Fig. 6}}). In the discussed application case, these would be areas like \textit{Oncology}, \textit{Immunology}, \textit{Indications}, \textit{Biomarkers}, \textit{Modalities}, \textit{Market Intelligence} or others. In the map analogy, these areas represent the “continents”. By zooming in to a lower elevation using an altitude slider or the mouse wheel, subdomains start to appear, until on highest granularity individual the data points, in the described application case scientific publications, become visible. When an individual entity is selected an information panel opens, providing bibliographical details of the selected paper including a link to the full text, while at the same time connected clones are highlighted on the map, enabling the understanding of semantic embeddings.

\begin{figure}[!ht]
    \centering
    \includegraphics[width=1\linewidth]{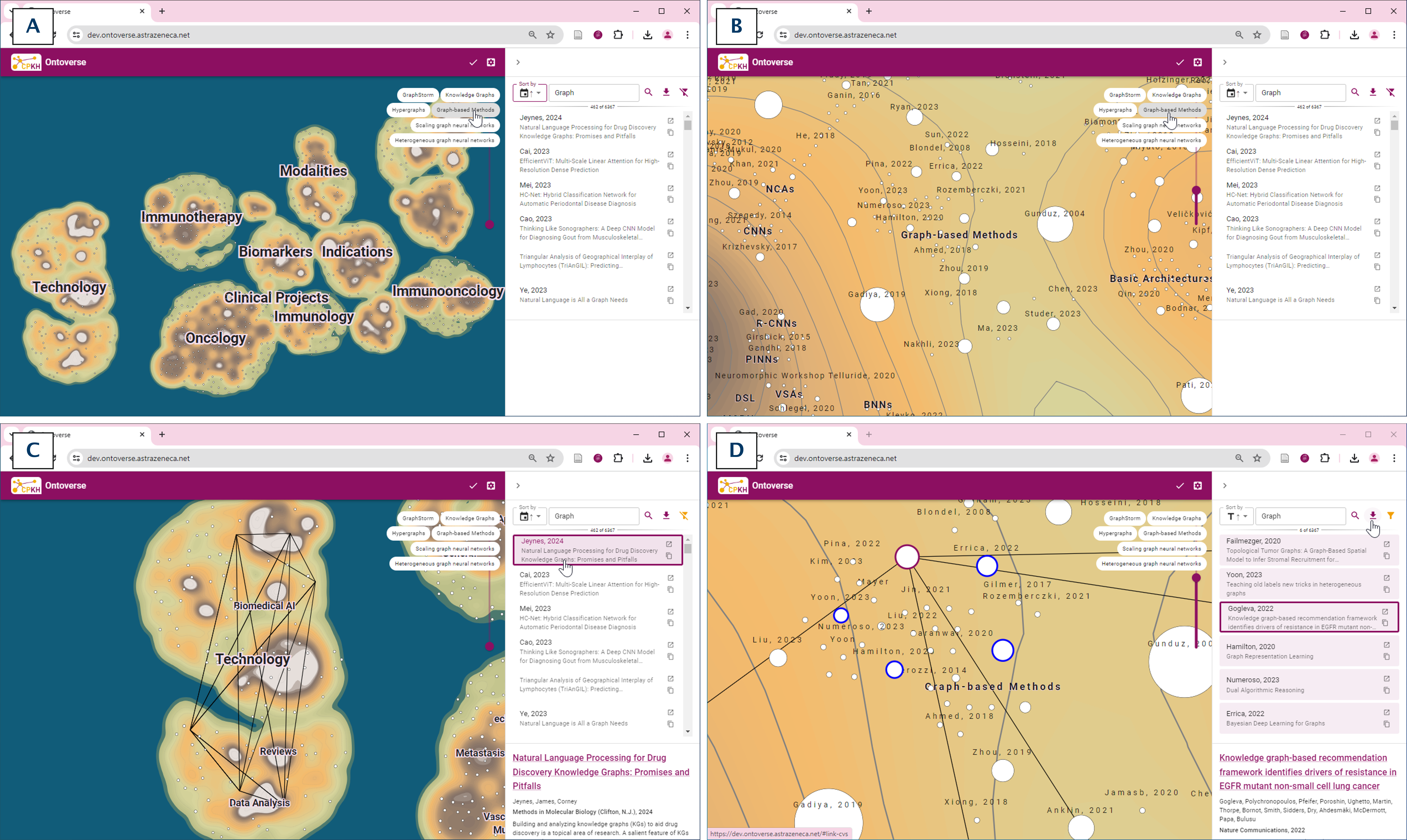}
    \caption{{\bf{Directed Navigation}}. Topics as well as papers can be queried in the search field (A). Selecting a topic takes the user to the respective region in the map (B). Selecting a paper from the list displays all clonal instances of the respective paper and their connections across topics (C). Results lists for export can be compiled by either multiply selecting nodes in the map or papers in the list (D).}
    \label{fig7:directed-navigation}
\end{figure}

\textbf{Directed navigation: } Alternatively, topics can be directly searched in a map search window, as in standard cartographic interfaces (\hyperref[fig7:directed-navigation]{\textbf{Fig. 7}}). Results for domains appear as topic proposal boxes at the edge of the map, while at the same time results for individual papers corresponding to the search criteria are shown in a list. Hitting a topic box takes the user to the respective region on the map. For categorically higher order topics the window will display a high-altitude view; for very granular topics the perspective will be zoomed in close to ground level. Similarly to the topic box display, the paper result list is also interactive. Selecting a list item will display the respective bibliographical information in the information window, and every clone is displayed in its respective domain along with thr connecting edges. This functionality allows for the rapid exploration of connections that are not as obvious and hidden relationships. It also invites for the discovery of adjacent publications and fields of research over the continuum of categorical levels. If an item is selected and, therefore, highlighted in the list, all topics to which it is assigned are highlighted as boxes in the information window. 
\textbf{Sharing of results:} Search results can be ranked and ordered according to the usual criteria, including author, title, year of publication, and exported as records to generate reference lists. Bibliographical lists can be compiled by either selecting papers directly in the map, or from the list, and be exported in tabular format. Regions in the map can easily be exchanged between co-workers via sharing the URL of the present viewing frame. The shared URL encodes the geographical coordinates of viewing frames at a respective zoom level as longitude, latitude and altitude which facilitates a straightforward communication of areas of interest instead of tedious sending back and forth between individual publications.

\section*{Discussion and Future Directions}
As reigning in the complexity of the exploding volume of biomedical knowledge is a generally acknowledged necessity, there is a burgeoning number of approaches, both in the academic and industry realms, enabling the fluent navigation of bodies of data. Two main concepts have preferentially been investigated: Exploration via a \textit{dynamic} graph requiring (and starting from) a specific seed point, or depiction of knowledge as a \textit{fixed} landscape. Early non-profit initiatives to provide seed point based tools to the community \cite{EigenFactor2024},\cite{krakerOpenKnowledgeMaps2020} were followed by commercially available a one-shot interfaces \cite{eitanConnectedPapers2024}, multi-seed iterative approaches \cite{Inciteful2024},\cite{LitMaps2024} or graph-centric visualizations \cite{ZetaAlpha2024},\cite{Semspect2024}. None of these approaches satisfied our specific needs, as they could not combine the requirements of input data agnosticism, conceptual holism, maximally intuitive usage concept and multi-topic occupancy. More importantly, the clustering and connectivity between papers commonly relies on algorithmic approaches that do no explicitly take into account precise mapping to domain knowledge, but rather statistical measures or embeddings that amount to a distribution over topic space that may not reflect reality.

We have implemented a highly customisable, generic framework to structure information by radically deconvoluting the complexity of Knowledge Graph-based data, mapping information into a landscape following strictly a cartographic metaphor and an ubiquitously adopted navigation paradigm, unlocking data exploration, analysis, and discovery for a broad target audience. Hierarchical topic modelling uses paper annotations as keywords to automatically assign papers to topics and to build the THG. A novel method allows for multi-topic occupancy whereby individual papers can appear in more than one location in the map so that we can visualise connections between topics and the contribution of papers across multiple domains. Our topographical paradigm and its underlying architecture for structuring and navigating the underlying information structures has the potential for future adoption of other graph analytic and machine learning approaches to deliver greater functionality and improve the end-use experience.

As next steps we envision the ingestion of items via the interface directly users, enabling a crowdsourcing of the KG and subsequently the map. 
In addition, to further leverage the enormous potential of pairing NLP with KGs \cite{Pan_Luo_Wang_Chen_Wang_Wu_2023} also on the UX end, we plan to integrate chat functionality with the interface to maximise the efficiency of the insights generation process for the user through free-text interrogation of the landscape.


The rapid expansion in the capabilities of Natural Language Processing (NLP) techniques including Large Language Models (LLMs) marks a significant step-change in our ability to derive knowledge from unstructured textual data. These advanced methods extend NLP capabilities far beyond traditional methods, particularly in Named Entity Recognition (NER and Named Entity Normalisation (NEN). By understanding the nuances and varied expressions used by different researchers more comprehensively these methods allow for enhanced precision and depth of understanding. These methods can be integrated with internal and external ontologies providing flexible, but explicit, mapping into the scientific community's vast exiting structured knowledge space. Combining these developments with Neo4j's graph infrastructure including it's powerful vector index capabilities enables semantic search at scale and critically, the retrieval of contextually relevant information based on meaning rather than just keywords.

Combining the resulting knowledge graph with LLMs and refining their ability to leverage information from the native knowledge graph and associated ontologies promises exciting new future functionalities. Task refinement and context adaptation can be achieved through techniques such as retrieval-augmented generation (RAG) and expansion of rich, user-friendly interaction via chatbots and textual question answering features. Grounding our models against existing structured domain knowledge using ontologies will allow us to tackle the potential for hallucination effects, for example by using evidence retrieval and validation approaches and RAG. The end goal is to build reliable, fact-based conversational agents that enhance user interaction and provide accurate, context-aware responses \cite{Agrawal_Kumarage_Alghamdi_Liu_2024}. Moreover, the introduction of chatbot interfaces democratises conversational access to available data, allowing more researchers to query knowledge without needing prior expertise in programming languages. This opens the door for a broader range of researchers to leverage advanced data insights, fostering innovation and accelerating target identification and drug discovery in general.

Lastly, the \textit{ab initio} generation of Ontoverse maps directly in the interface in an integrated, user-centric approach, with the nurturing of data into the system being guided by the combination of an iterative scraping-, recommendation-, and acceptance-rejection process, is a logical next step in the evolution of the concept. This will further drive the democratisation of knowledge aggregation and dissemination for domain-specific user communities.


\bibliographystyle{unsrt}
\bibliography{ontoverse}

\end{document}